\documentclass[12pt]{article}

\usepackage{amssymb,amsthm}
\usepackage{graphicx}
\usepackage{fancyref}
\usepackage{multirow}
\usepackage{array}
\usepackage{amsmath,amsfonts}
\usepackage{longtable}
\usepackage{bm}
\usepackage{lmodern}
\usepackage{epsfig}
\usepackage{indentfirst}
\usepackage{url}
\usepackage{hyperref}
\renewcommand*\&{and}

\usepackage{overpic}
\usepackage{supertabular}
\usepackage{array}
\usepackage[super,square,sort&compress]{natbib}
\makeatletter                            
\renewcommand\@biblabel[1]{#1.}          

\usepackage{float}    
\usepackage{setspace}
\usepackage{enumerate}  
\usepackage{epstopdf}
\usepackage{rotating} 
\usepackage{lscape}  
\usepackage{extarrows}    
\usepackage{textcomp}

\usepackage[latin1]{inputenc}
\usepackage{tikz}
\usetikzlibrary{shapes,arrows,chains}
\usetikzlibrary{decorations.pathmorphing, backgrounds, positioning, fit, petri, automata}
\usepackage{xcolor}

\definecolor{yellow1}{rgb}{1,0.8,0.2}      
\definecolor{LightBlue1}{RGB}{202,225,255}         
\definecolor{SteelBlue3}{RGB}{79,148,205}

\newtheorem{lemma}{Lemma}
\newtheorem{theorem}{Theorem}

\theoremstyle{definition}

\begin{document}
\title{{Optimally estimating the sample standard deviation from the five-number summary}}

{\footnotesize
\author{\small Jiandong Shi$^{1}, $ Dehui Luo$^{1}$, Hong Weng$^{2}, $ Xian-Tao Zeng$^{2}, $ Lu Lin$^{3,4}, $\\
\small Haitao Chu$^{5}$ and Tiejun Tong$^{1,}$\thanks{Corresponding author. E-mail: tongt@hkbu.edu.hk} \\ \\
{\footnotesize  $^1$Department of Mathematics, Hong Kong Baptist University, Hong Kong}\\
{\footnotesize  $^2$Center for Evidence-Based and Translational Medicine, Zhongnan Hospital of}\\
{\footnotesize Wuhan University, Wuhan, China}\\
{\footnotesize $^3$School of Statistics, Shandong Technology and Business University, Yantai, China}\\
{\footnotesize $^4$Zhongtai Securities Institute for Finance Studies, Shandong University, Jinan, China}\\
{\footnotesize $^5$Division of Biostatistics, School of Public Health, University of Minnesota, Minneapolis, USA}\\
}
}

\date{}
\maketitle

\begin{abstract}
When reporting the results of clinical studies, some researchers may choose the five-number summary (including the sample median, the first and third quartiles,
and the minimum and maximum values) rather than the sample mean and standard deviation, particularly for skewed data.
For these studies, when included in a meta-analysis,
it is often desired to convert the five-number summary back to the sample mean and standard deviation.
For this purpose, several methods have been proposed in the recent literature and they are increasingly used nowadays.
In this paper, we propose to further advance the literature
by developing a smoothly weighted estimator for the sample standard deviation
that fully utilizes the sample size information.
For ease of implementation, we also derive an approximation formula for the optimal weight,
as well as a shortcut formula for the sample standard deviation.
Numerical results show that our new estimator provides a more accurate estimate
for normal data and also performs favorably for non-normal data.
Together with the optimal sample mean estimator in Luo et al.,\cite{luo2018optimally}
our new methods have dramatically improved the existing methods for data transformation, and they are capable to serve as ``rules of thumb" in meta-analysis for studies reported with the five-number summary.
Finally for practical use, an Excel spreadsheet and an online calculator are also provided
for implementing our optimal estimators.

\

\noindent
$Key \ words$:
Five-number summary,
Interquartile range,
Range,
Sample mean,
Sample size,
Standard deviation
\end{abstract}

\newpage
\baselineskip 22pt
\setlength{\parskip}{0.2\baselineskip}
\section{Introduction}
\noindent
Meta-analysis is becoming increasingly popular in the past several decades, mainly owing to its wide range of applications in evidence-based medicine.\citealp{evidence1992evidence}$^-$\citealp{borenstein2009introduction}
To statistically combine data from multiple independent studies,
researchers need to first conduct a systematic review and extract the summary data
from the clinical studies in the literature.
For continuous outcomes, e.g., the blood pressure level and the amount of alcohol consumed,
the sample mean and standard deviation (SD) are the most commonly used summary statistics
for evaluating the effectiveness of a certain medicine or treatment.
For skewed data, however, the five-number summary
(including the  sample median, the first and third quartiles, and the minimum and maximum values)
has also been frequently reported in the literature.
To the best of our knowledge, there are few methods available in meta-analysis
that can incorporate the sample median and the sample mean simultaneously.
As an example, when applying the fixed-effect model or the random-effects model, the sample mean and SD are the must-to-have quantities for computing the overall effect.\cite{higgins2008cochrane}

This then yields a natural question as follows: when performing a meta-analysis,
how to deal with clinical studies in which the five-number summary was reported
rather than the sample mean and SD?
In the early stages, researchers often exclude such studies from further analysis
by claiming them as ``studies with no sufficient data" in the flow chart of study selection.
Such an approach is, however, often suboptimal as it excludes valuable information from the literature.
And consequently, the final results are less reliable,
in particular when a large proportion of the included studies are with the five-number summary.
For this, there is an increased demand for developing new methods that are able to
convert the five-number summary back to the sample mean and SD.
For ease of notation, let $\{a,q_1,m,q_3,b\}$ denotes the five-number summary, where $a$ is the sample minimum,
$q_1$ is the first quartile, $m$ is the sample median, $q_3$ is the third quartile, and $b$ is the sample maximum of the data.
We also let $n$ be the sample size in the study.

Note that the five-number summary may not be fully reported in clinical studies.
In a special case with $\{a,m,b\}$ being reported,
Hozo et al.\cite{hozo2005estimating} was among the first to estimate the sample mean and SD.
It is noted, however, that Hozo et al.\cite{hozo2005estimating} did not sufficiently use the information of sample
size $n$ so that their estimators are either biased or non-smooth.
Inspired by this, Wan et al.\cite{wan2014estimating} and Luo et al.\cite{luo2018optimally} further improved the existing methods
by proposing nearly unbiased and optimal estimators with analytical formulas.
In addition, we note that Walter and Yao \cite{walter2007effect} had also provided a numerical solution
for estimating the sample SD,
while the lack of the analytical formula makes it less accessible to practitioners.
In another special case with $\{q_1, m, q_3\}$ being reported,
Wan et al.\cite{wan2014estimating} proposed a nearly unbiased estimator for the sample SD, and
Luo et al.\cite{luo2018optimally} proposed an optimal estimator for the sample mean
by fully using the sample size information.
In Google Scholar as of 27 February 2020,
Hozo et al.,\cite{hozo2005estimating} Wan et al.\cite{wan2014estimating} and Luo et al.\cite{luo2018optimally}
have been cited 3595, 1078 and 143 times, respectively.
Without any doubt, these several papers have been attracting more attentions and playing an important role in meta-analysis.

When $\{a,q_1,m,q_3,b\}$ was fully reported, Bland \cite{Bland2015Estimating} extended Hozo et al.'s method
to estimate the sample mean and SD from the five-number summary.
As their methods are essentially the same, it is noted that the estimators in Bland \cite{Bland2015Estimating} are also suboptimal mainly because the sample size information is again not sufficiently used.
To be more specific, the sample mean estimator in Bland \cite{Bland2015Estimating} is
\begin{equation*}
  \bar{X}\approx\dfrac{a+2q_1+2m+2q_3+b}{8}=\dfrac{1}{4}\left(\dfrac{a+b}{2}\right)+\dfrac{1}{2}\left(\dfrac{q_1+q_3}{2}\right)+\dfrac{m}{4}.
\end{equation*}
According to Johnson \& Kuby, \cite{johnson2007elementary}
given that the data follow a symmetric distribution,
the quantities $(a+b)/2$, $(q_1+q_3)/2$, and $m$ can each serve as an estimate of the sample mean.
To have a final estimator, Bland \cite{Bland2015Estimating} applied the artificial weights 1/4, 1/2, and 1/4 for the three components, respectively.
That is, the first and third components are treated equally and both of them are only half reliable compared to the second component.
As this is not always the truth, to improve the sample mean estimation,
Luo et al.\cite{luo2018optimally} proposed the optimal estimator as
\begin{equation}
\bar{X}\approx w_{3,1}\left(\dfrac{a+b}{2}\right)+w_{3,2}\left(\dfrac{q_1+q_3}{2}\right)+(1-w_{3,1}-w_{3,2})m,
\label{mabq1q3}
\end{equation} 
\noindent where $w_{3,1}=2.2/(2.2+n^{0.75})$ and $w_{3,2}=0.7-0.72/n^{0.55}$ are the optimal weights
assigned to the respective components.
Specifically, they first derived the optimal weights
that minimize the mean squared error (MSE) of estimator (\ref{mabq1q3}).
Note however that the analytical forms of the optimal weights,
as shown in their formula (13),
are rather complicated and may not be readily accessible to practitioners.
They further derived the above $w_{3,1}$ and $w_{3,2}$ as the approximated optimal weights for practical use.

For the sample SD estimation from the five-number summary,
Bland \cite{Bland2015Estimating} also provided an estimator that follows the inequality method as in Hozo et al.\cite{hozo2005estimating}
Then Wan et al.\cite{wan2014estimating} proposed a nearly unbiased estimator of the sample SD to improve the literature.
In Section 3, we will point out that the sample SD estimator in Wan et al.\cite{wan2014estimating} may still not be optimal due to the insufficient use of the sample size information.
For more details, see the motivating example in Section 3.
According to Higgins \& Green \cite{higgins2008cochrane} and Chen \& Peace,\cite{chen2013applied} the sample SD plays a crucial role in weighting the studies in meta-analysis.
Inaccurate weighting results may lead to biased overall effect sizes and biased confidence intervals, and hence mislead physicians to provide patients with unreasonable or even wrong medications.
Inspired by this, we propose a smoothly weighted estimator for the sample SD to further improve the existing literature.
To promote the practical use, we have provided an Excel spreadsheet to implement the optimal estimators
in the Supplementary Material.
More importantly, we have also incorporated the new estimator in this paper into our online calculator
at \url{http://www.math.hkbu.edu.hk/~tongt/papers/median2mean.html}.
From the practical point of view, our proposed method will make a solid contribution to meta-analysis
and has the potential to be widely used.

The rest of the paper is organized as follows.
In Section 2, we review the existing methods for estimating the sample SD under three common scenarios.
In Section 3, we present a motivating example,
propose a smoothly weighted estimator for the sample SD,
and derive a shortcut formula of our new estimator for practical use.
In Section 4, we conduct numerical studies to assess the finite sample performance of our new estimator,
and meanwhile we demonstrate its superiority over the existing methods.
We then conclude the paper in Section 5 with a brief summary of the existing methods, and provide the theoretical results in Section 6.

\vskip 34.45pt
\section{Existing methods}
\noindent
Needless to say, the sample size $n$ provides an important information and should be sufficiently used in the estimation procedure.
To incorporate it with the five-number summary, we let $\mathcal{S}_1=\{a, m, b; n\}$, $\mathcal{S}_2=\{q_1, m, q_3; n\}$,
and $\mathcal{S}_3=\{a, q_1, m, q_3, b; n\}$, that represent the three common scenarios in the literature.
Clearly, $\mathcal{S}_1$ and $\mathcal{S}_2$ are two special cases of $\mathcal{S}_3$.
To further clarify, we also note that $\mathcal{S}_1$, $\mathcal{S}_2$ and $\mathcal{S}_3$ are the same as $\mathcal{C}_1$, $\mathcal{C}_3$ and $\mathcal{C}_2$ in Wan et al.,\cite{wan2014estimating} respectively.
In this section, we briefly review the existing estimators of the sample SD under the three scenarios.

\

\subsection{Estimating the sample SD from $\mathcal{S}_1 = \{a, m, b; n\}$}
\noindent
Under scenario $\mathcal{S}_1$, Hozo et al.\cite{hozo2005estimating} proposed to estimate the sample SD by
$$
S\approx\left\{
\begin{aligned}
&  \dfrac{1}{\sqrt{12}}\left[(b-a)^2+\dfrac{(a-2m+b)^2}{4}\right]^{1/2} \quad & n\leq15,\\
& \dfrac{b-a}{4}  & 15<n\leq70,\\
& \dfrac{b-a}{6}  &n>70.
\end{aligned}
\right.
$$
\noindent As shown in Google Scholar, Hozo et al.'s estimator is very popular in the previous literature due to the huge demand of data transformation in evidence-based medicine.
We note, however, that their estimator is a step function of the sample size $n$ so that the final estimate may not be optimal.
For example, when $n$ increases from 70 to 71, there is a sudden drop in the estimated value of the sample SD,
namely dropped by 33.3\% if the 71st sample is not an  extreme value so that $a$ and $b$ remain the same in both samples.
On the other hand, the sample size information is completely ignored within each of the three intervals so that their estimator is biased and suboptimal.

In view of the above limitations, Wan et al.\cite{wan2014estimating} proposed a nearly unbiased estimator of the sample SD as
\begin{equation}
  S\approx\dfrac{b-a}{\xi},
  \label{ab}
\end{equation}
where $\xi=\xi(n)=2\Phi^{-1}[(n-0.375)/(n+0.25)]$, $\Phi$ is the cumulative distribution function of the standard normal distribution, and $\Phi^{-1}$ is the inverse function of $\Phi$.
Wan et al.'s estimator not only overcomes the limitations of Hozo et al.'s estimator by incorporating the sample size information efficiently, but also is very simple in practice for the given analytical formula.
From a statistical point of view, Wan et al.\cite{wan2014estimating} has provided the best method for estimating the sample SD,
given that the reported data include only the four numbers in scenario $\mathcal{S}_1$.

\

\subsection{Estimating the sample SD from $\mathcal{S}_2 = \{q_1, m, q_3; n\}$}
\noindent
Under scenario $\mathcal{S}_2$, Wan et al.\cite{wan2014estimating} proposed an estimator of the sample SD as
\begin{equation}
  S\approx\dfrac{q_3-q_1}{\eta},
  \label{q1q3}
\end{equation}
where $\eta=\eta(n)=2\Phi^{-1}[(0.75n-0.125)/(n+0.25)]$.
Note that the method for deriving estimator (\ref{q1q3}) is the same in spirit as that for estimator (\ref{ab}).
In particular, Wan et al.\cite{wan2014estimating} has incorporated the sample size information appropriately so that the proposed estimator is nearly unbiased.
When the reported data include only $q_1$, $m$, $q_3$ and $n$, it can be shown that estimator (\ref{q1q3}) is the best method for estimating the sample SD.

\

\subsection{Estimating the sample SD from $\mathcal{S}_3 = \{a, q_1, m, q_3, b; n\}$}
\noindent
Under scenario $\mathcal{S}_3$, Bland \cite{Bland2015Estimating} proposed to estimate the sample SD by
\begin{equation*}
\begin{split}
 S \approx &  \ \left[{(a^2+2q_1^2+2m^2+2q_3^2+b^2) \over 16}+{(aq_1+q_1m+mq_3+q_3b)\over 8} \right.\\
           &  \ \left. -{(a+2q_1+2m+2q_3+b)^2\over 64} \large\right]^{1/2}.
\end{split}
\end{equation*}
\noindent 
As mentioned in Section 1, Bland's estimator is independent of the sample size and is less accurate for practical use.
To improve the literature, Wan et al.\cite{wan2014estimating} proposed the following estimator of the sample SD:
\begin{equation}\label{abq}
  S\approx\dfrac{1}{2}\left(\dfrac{b-a}{\xi}+\dfrac{q_3-q_1}{\eta}\right).
\end{equation}

In essence, Wan et al.\cite{wan2014estimating} treated scenario $\mathcal{S}_3$ as a combination of scenario $\mathcal{S}_1$ and scenario $\mathcal{S}_2$.
By estimating the sample SD from scenario $\mathcal{S}_1$ and scenario $\mathcal{S}_2$ separately, they applied the average of estimators (\ref{ab}) and (\ref{q1q3}) as the final estimator.
In Section 3, we will show that $(b-a)/\xi$ and $(q_3-q_1)/\eta$ may not be equally reliable for different sample size $n$.
As a consequence, the average estimator in formula (\ref{abq}) is not the optimal estimator due to the insufficient use of the sample size information.
This motivates us to propose a smoothly weighted estimator for the sample SD to further improve Wan et al.'s estimator in this paper.

\

\section{Main results}
\noindent
To present the main idea, we let $X_{1} , X_{2} , \ldots , X_{n}$ be a random sample of size $n$ from the normal distribution with mean $\mu$ and variance $\sigma^2$,
and $X_{(1)} \leq X_{(2)} \leq \cdots \leq X_{(n)}$ be the order statistics of $X_{1} , X_{2} , \ldots , X_{n}$.
Let also $X_i = \mu+\sigma Z_i$ for $i=1, 2, \ldots,n$.
Then $Z_{1} , Z_{2} , \ldots , Z_{n}$ follow the standard normal distribution with $Z_{(1)}\leq Z_{(2)} \leq \cdots \leq Z_{(n)}$ as the corresponding order statistics.
Finally, by letting $n = 4Q+1$ with $Q$ a positive integer, we have $a = \mu+\sigma Z_{(1)}$, $q_1 = \mu+\sigma Z_{(Q+1)}$, $m = \mu+\sigma Z_{(2Q+1)}$, $q_3 = \mu+\sigma Z_{(3Q+1)}$, and $b = \mu+\sigma Z_{(n)}$.

\

\subsection{Motivating example}
\begin{figure}[H]
  \centering
  \includegraphics[width=1\textwidth]{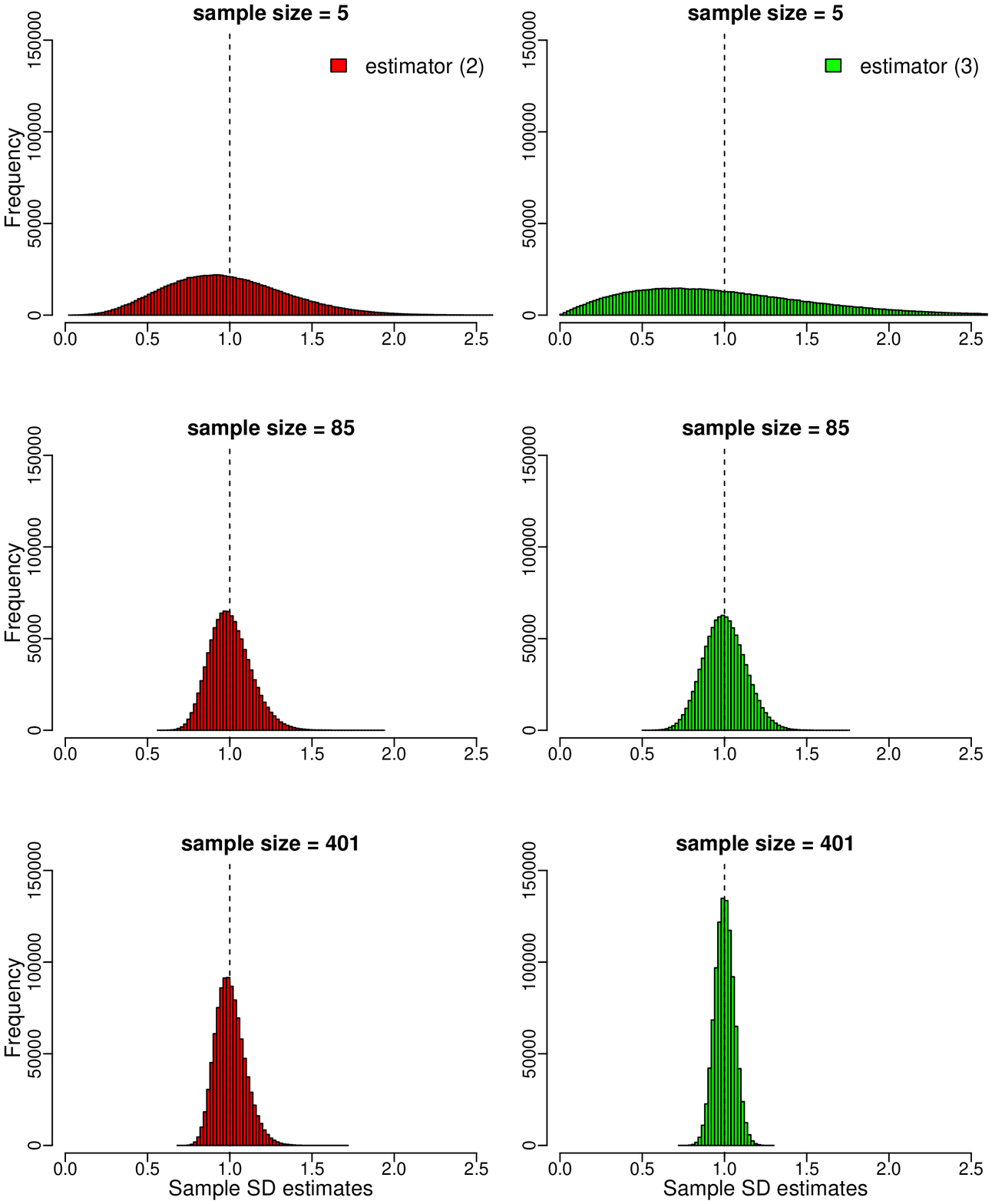}
  \caption{The histograms of the sample SD estimates (the true SD is 1 as shown in the vertical dashed lines)
  with the sample sizes 5, 85 and 401, with a total of 1,000,000 simulations.
  The red and green histograms represent the frequencies of the estimates
  by estimators (\ref{ab}) and (\ref{q1q3}), respectively.
  }\label{fig:motivation}
\end{figure}

\noindent
To investigate whether the two components in estimator (\ref{abq}) are equally reliable, we first conduct a simple simulation study.
In each simulation, we generate a random sample of size $n$ from the standard normal distribution, find the five-number summary $\{a,q_1,m,q_3,b\}$,
and then apply estimators (\ref{ab}) and (\ref{q1q3}) to estimate the sample SD respectively.
For $Q=1,21\ {\rm and} \ 100$, or equivalently, $n=5, 85\ {\rm and}\ 401$,
we repeat the simulation 1,000,000 times and plot the histograms of the sample SD estimates
in Figure \ref{fig:motivation} for both methods.


From Figure \ref{fig:motivation}, it is evident
that estimators (\ref{ab}) and (\ref{q1q3}) may not be equally reliable when the sample size varies.
Note that the true SD equals one since the data are generated from the standard normal distribution.
When the sample size is small (say $n=5$),
estimator (\ref{ab}) provides a more accurate and less skewed estimate for the sample SD.
When the sample size is moderate (say $n=85$),
the two estimators perform similarly and are about equally reliable.
When the sample size is large (say $n=401$),
estimator (\ref{q1q3}) provides to be a more reliable estimator than estimator (\ref{ab}).
This hence shows that Wan et al.'s estimator in formula (\ref{abq}) may not be the optimal estimator
for the sample SD when the five-number summary is fully reported.
We propose to further improve Wan et al.'s estimator
by considering a linear combination of estimators (\ref{ab}) and (\ref{q1q3}),
in which the optimal weight is a function of the sample size.

\

\subsection{Optimal sample SD estimation}
\noindent
In view of the limitations of estimator (\ref{abq}), we propose the following estimator for the sample SD:
\begin{equation}\label{abq1q3}
S_w=w\left(\dfrac{b-a}{\xi}\right)+(1-w)\left(\dfrac{q_{3}-q_{1}}{\eta}\right),
\end{equation}
where $\xi$ and $\eta$ are given in Section 2,
and $w$ is the weight assigned to the first component.
Note that the new estimator is a weighted combination of estimators (\ref{ab}) and (\ref{q1q3}).
When $w=1$, the new estimator reduces to estimator (\ref{ab}).
When $w=0$, the new estimator reduces to estimator (\ref{q1q3}).
And when $w=0.5$, the new estimator leads to estimator (\ref{abq}) in Wan et al.\cite{wan2014estimating}
Hence, the existing estimators of the sample SD are all special cases of our new estimator.

To find the optimal estimator, we consider the commonly used quadratic loss function, i.e., $L(S_w, \sigma) = (S_w-\sigma)^2$.
We select the optimal weight by minimizing the expected value of the loss function $L(S_w,\sigma)$, or equivalently, by minimizing the mean squared error (MSE) of the estimator.
In Section 6, we show in Theorem \ref{theorem1} that the optimal weight is, approximately,
\begin{equation}\label{weight}
w_{\rm opt}= \dfrac{{\rm Var}(q_{3}-q_{1})/\eta^{2}-{\rm Cov}(b-a, q_{3}-q_{1})/(\xi\eta)}{{\rm Var}(b-a)/\xi^{2}+{\rm Var}(q_{3}-q_{1})/\eta^{2}-2{\rm Cov}(b-a,q_{3}-q_{1})/(\xi\eta)}.
\end{equation}
We further show that the optimal weight will converge to zero when the sample size tends to infinity.
That is, estimator (\ref{q1q3}) will be more reliable than estimator (\ref{ab}) when the sample size is large.

Note that the optimal weight in formula (\ref{weight}) has a complicated form
and may not be readily accessible to practitioners.
To promote the practical use of the new estimator,
we also develop an approximation formula for the optimal weight.
Recall that $a = \mu+\sigma Z_{(1)}$, $q_1 = \mu+\sigma Z_{(Q+1)}$, $q_3 = \mu+\sigma Z_{(3Q+1)}$, and $b = \mu+\sigma Z_{(n)}$.
We have $b-a=\sigma(Z_{(n)}-Z_{(1)})$ and $q_3-q_1 = \sigma(Z_{(3Q+1)}-Z_{(Q+1)})$.
Then by formula (\ref{weight}) and the symmetry of the standard normal distribution, we can rewrite the optimal weight as
\begin{equation}\label{approximation}
w_{\rm opt}(n) = \dfrac{1}{1+J(n)},
\end{equation}
where
\begin{equation}\label{Jn}
J(n)=\dfrac{{\rm Var}(Z_{(n)}-Z_{(1)})/\xi^2-{\rm Cov}(Z_{(n)}-Z_{(1)}, Z_{(3Q+1)}-Z_{(Q+1)})/(\xi\eta)}{{\rm Var}(Z_{(3Q+1)}-Z_{(Q+1)})/\eta^2-{\rm Cov}(Z_{(n)}-Z_{(1)}, Z_{(3Q+1)}-Z_{(Q+1)})/(\xi\eta)}.
\end{equation}

\noindent
Note that $Z_{(1)}$, $Z_{(Q+1)}$, $Z_{(3Q+1)}$ and $Z_{(n)}$
are the order statistics of the standard normal distribution,
and therefore their variances and covariances do not depend on $\mu$ and $\sigma^2$.
In addition, by formulas (\ref{ab}) and (\ref{q1q3}),
$\xi$ and $\eta$ are both the functions of the sample size $n$ only.
This shows that $J(n)$ is independent of $\mu$ and $\sigma^2$,
and consequently, the optimal weight $w_{\rm opt}$ is also a function of $n$ only.
For clarification, we have expressed the optimal weight as $w_{\rm opt} = w_{\rm opt}(n)$ in formula (\ref{approximation}).

\

\subsection{An approximation formula}
\begin{figure}[H]
\setlength{\abovecaptionskip}{6pt}%
\setlength{\belowcaptionskip}{0pt}%
  \centering
  \vspace{-0.2in}
  \includegraphics[width=1\textwidth]{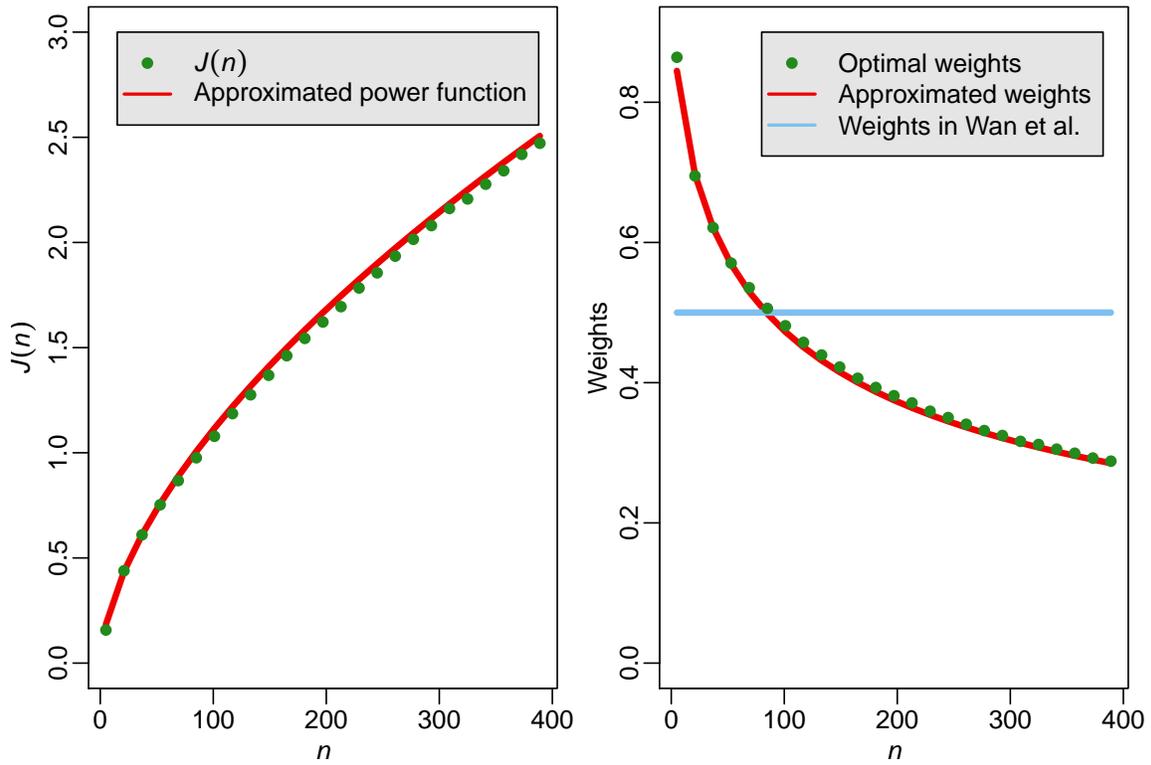}
  \caption{The left panel displays the true and approximated values of $J(n)$,
  and the right panel displays the optimal weights, the approximated weights and the weights in Wan et al.\cite{wan2014estimating}}
\label{fig:optimalweight}
\end{figure}

\noindent
To have an approximation formula for the optimal weight, we numerically compute the true values of $J(n)$ and $w_{\rm opt}(n)$ for different values of $n$ using formulas (\ref{approximation}) and (\ref{Jn}).
We then plot $J(n)$ and $w_{\rm opt}(n)$ in Figure \ref{fig:optimalweight}
for $n$ varying from 5 to 401, respectively.
Observing that $J(n)$ is an increasing and concave function of $n$, we consider the simple power function $c_1n^{c_2}+c_0$ to approximate $J(n)$ with $0<c_2<1$ so that the approximation curve is also concave.
With the true values of $J(n)$, the best values of the coefficients are approximately $c_1=0.07$, $c_2=0.6$ and $c_0=0$.
Finally, by plugging $J(n)\approx0.07n^{0.6}$ into formula (\ref{approximation}), we have the approximation formula for the optimal weight as
\begin{equation}\label{aweight}
\tilde{w}_{\rm opt}(n)\approx \dfrac{1}{1+0.07n^{0.6}}.
\end{equation}

From the right panel of Figure \ref{fig:optimalweight}, it is evident that the approximation formula provides a perfect fit to the true optimal weight values for $n$ up to $401$.
By formulas (\ref{abq1q3}) and (\ref{aweight}), our proposed estimator of the sample SD from the five-number summary is
\begin{equation}\label{approx}
S(n)\approx\left(\dfrac{1}{1+0.07n^{0.6}}\right)\dfrac{b-a}{\xi}+\left(\dfrac{0.07n^{0.6}}{1+0.07n^{0.6}}\right)\dfrac{q_3-q_1}{\eta}.
\end{equation}

\noindent 
By formula (\ref{aweight}) and the fact that $n=4Q+1$,
we note that the nearest integer for $Q$ such that $\tilde{w}_{\rm opt}(n)\approx0.5$ is $Q=21$.
In other words, estimators (\ref{ab}) and (\ref{q1q3}) will be about equally reliable when $n=85$.
This coincides with our simulation results in Figure \ref{fig:motivation}
that estimators (\ref{ab}) and (\ref{q1q3}) perform very similarly when $n=85$.
This, from another perspective, demonstrates that our approximation formula can serve as a ``rule of thumb'' for estimating the sample SD from the five-number summary.

Recall that $\xi=\xi(n)=2\Phi^{-1}[(n-0.375)/(n+0.25)]$ and $\eta=\eta(n)=2\Phi^{-1}[(0.75n-0.125)/(n+0.25)]$,
where $\Phi^{-1}(z)$ is the upper $z$th quantile of the standard normal distribution.
We have the shortcut formula of estimator (\ref{approx}) as
\begin{equation}\label{shortcut}
S\approx\dfrac{b-a}{\theta_1}+\dfrac{q_3-q_1}{\theta_2},
\end{equation}
\noindent where
\begin{align*}
\theta_1=\theta_1(n) &= (2+0.14n^{0.6})\cdot \Phi^{-1}\left(\frac{n-0.375}{n+0.25}\right),\\
\theta_2=\theta_2(n) &= \left(2+\frac{2}{0.07n^{0.6}}\right)\cdot \Phi^{-1}\left(\frac{0.75n-0.125}{n+0.25}\right).
\end{align*}
For ease of implementation, we also provide the numerical values of $\theta_1$ and $\theta_2$ in Table \ref{coefficients} for $Q$ up to 100, or equivalently, for $n$ up to 401.
For a general sample size, one may refer to our Excel spreadsheet for specific values in the Supplementary Material,
or compute them using the command ``qnorm($z$)" in the R software.

\section{Numerical studies}
\noindent
To evaluate the practical performance of the new method, we conduct numerical studies to compare our proposed estimator with the three estimators in Wan et al.\cite{wan2014estimating}
The robustness of the estimators will also be examined.

In the first study, we generate the data from the normal distribution with mean 50 and standard deviation 17,
for which we follow the same settings as in Hozo et al.\cite{hozo2005estimating} and Wan et al.\cite{wan2014estimating}
Then for the simulated data with sample size $n$, we compute the sample SD, denoted as $S^{\rm Sam}$,
and also record the five-number summary $\{a , q_1, m, q_3, b \}$.
To apply the proposed method, we assume that all the available data are the five-number summary and the sample size,
and apply estimators (\ref{ab}), (\ref{q1q3}), (\ref{abq}) and (\ref{shortcut}) to estimate the sample SD, denoted by $S_1$, $S_0$, $S_{0.5}$ and $S_{\tilde{w}_{\rm opt}}$, respectively.
Finally, for a fair comparison, we compute the relative mean squared errors (RMSE) of the four estimators as follows:
\begin{align*}
  {\rm RMSE}(S_1)=\dfrac{\sum_{i=1}^{T}(S_{1,i}-\sigma)^2}{\sum_{i=1}^{T}(S_i^{\rm Sam}-\sigma)^2},  ~~~ & ~~~
  {\rm RMSE}(S_0)=\dfrac{\sum_{i=1}^{T}(S_{0,i}-\sigma)^2}{\sum_{i=1}^{T}(S_i^{\rm Sam}-\sigma)^2}, \nonumber \\
    {\rm RMSE}(S_{0.5})=\dfrac{\sum_{i=1}^{T}(S_{0.5,i}-\sigma)^2}{\sum_{i=1}^{T}(S_i^{\rm Sam}-\sigma)^2},  ~~~ & ~~~  {\rm RMSE}(S_{\tilde{w}_{\rm opt}})=\dfrac{\sum_{i=1}^{T}(S_{\tilde{w}_{\rm opt},i}-\sigma)^2}{\sum_{i=1}^{T}(S_i^{\rm Sam}-\sigma)^2},
\end{align*}
where $T$ is the total number of simulations, $\sigma=17$ is the true SD,
and $S_i^{\rm Sam}$ is the sample SD in the $i$th simulation.

\begin{figure}[H]
  \centering
  \includegraphics[width=1\textwidth]{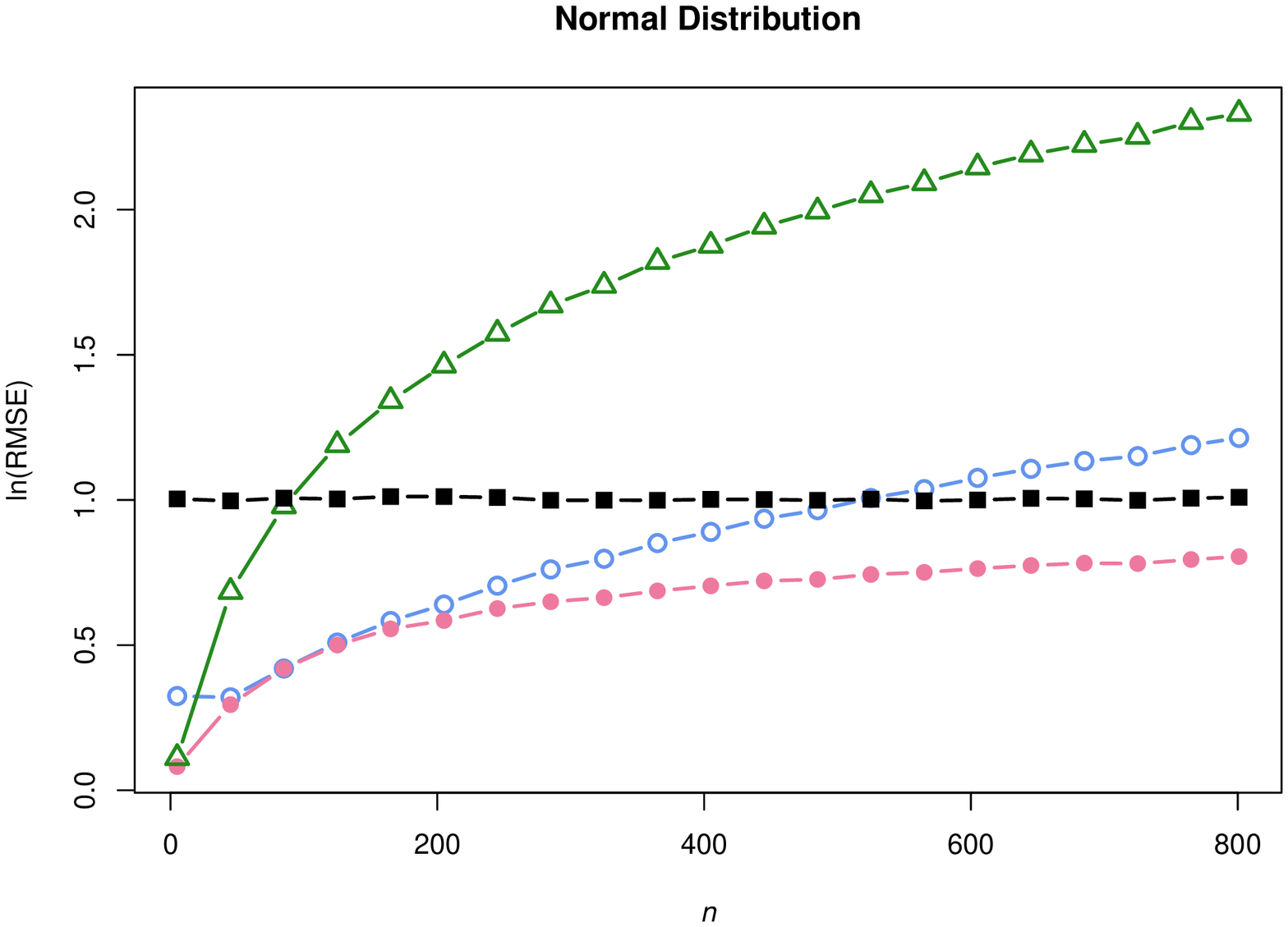}
  \caption{The natural logarithm of the RMSE values of the sample SD estimators for data from the normal distribution. The empty triangles represent the results by estimator (\ref{ab}), the solid squares represent the results by estimator (\ref{q1q3}), the empty circles represent the results by estimator (\ref{abq}), and the solid circles represent the results by our new estimator (\ref{shortcut}).}\label{fig:normal}
\end{figure}

With $T=2,000,000$ and $n$ ranging from 5 to 801, we compute the natural logarithm of the RMSE values for the four estimators and plot them in Figure \ref{fig:normal}.
From the numerical results, it is evident that our new estimator
has a smaller RMSE value than the three existing estimators in all settings,
which demonstrates that
our new estimator does provide the optimal estimate for the sample SD.
Specifically, for estimator (\ref{ab}) that only applies the minimum and maximum values,
it performs well only when the sample size is extremely small ($n<9$).
For estimator (\ref{q1q3}) that only applies the first and third quartiles,
it does not perform well for any sample size.
While for the equally weighted estimator (\ref{abq}),
we note that it performs better than estimators (\ref{ab}) and (\ref{q1q3}) in a wide range of settings.
Nevertheless, we also note that estimator (\ref{abq}) is not as good as estimator (\ref{ab})
when $n$ is relatively small ($n<21$), and is not as good as estimator (\ref{q1q3}) when $n$ is relatively large ($n>521$).
This shows, from another perspective, that estimator (\ref{abq}) does not provide an optimal weight
between the two elementary estimators and hence is still suboptimal.
In fact, compared to our new estimator, estimator (\ref{abq}) is capable to provide an optimal estimate only
when the sample size is about 85, which coincides with our analytical results in Section 3.3.
It is also interesting to point out that the numerical results for large sample sizes are also consistent with
the asymptotic results in Theorem \ref{Rmse},
in which we demonstrated that our new estimator has the smallest asymptotic RMSE among the four estimators.
To conclude, from both practical and theoretical perspectives,
our new estimator is superior to the existing estimators in all settings,
and it deserves as the optimal estimator of the sample SD for the studies reported with the five-number summary.

\begin{figure}[H]
  \centering
  \includegraphics[width=0.9\textwidth]{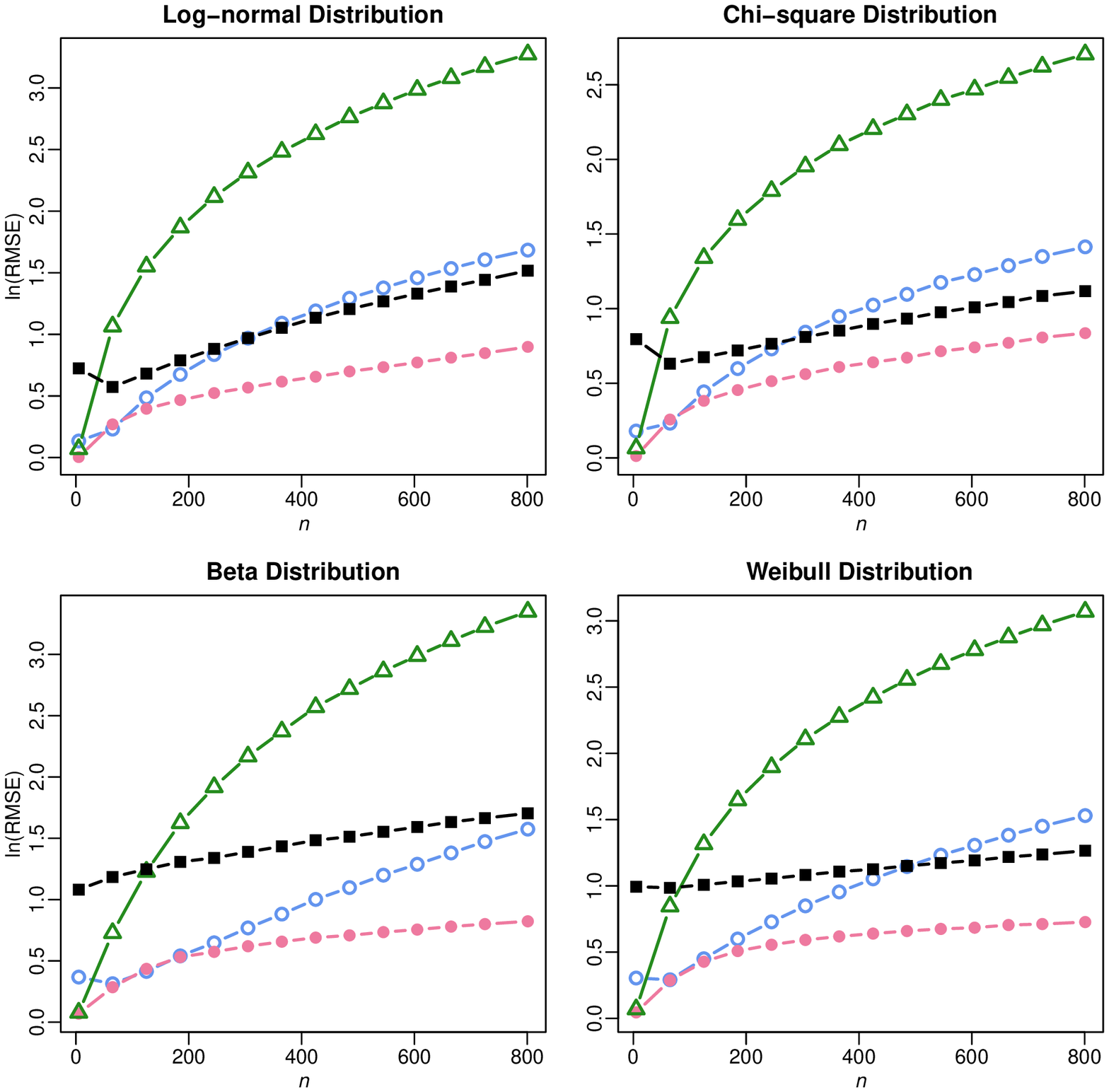}
  \caption{The natural logarithm of the RMSE values of the sample SD estimators for data from the skewed distributions. The empty triangles represent the results by estimator (\ref{ab}), the solid squares represent the results by estimator (\ref{q1q3}), the empty circles represent the results by estimator (\ref{abq}), and the solid circles represent the results by our new estimator (\ref{shortcut}).}\label{fig:distribution}
\end{figure}

To check the robustness of our proposed estimator,
we conduct another numerical study with data generated from non-normal distributions.
Specifically, we consider four skewed distributions including the log-normal distribution with location parameter $\mu=4$ and scale parameter $\sigma=0.3$, the chi-square distribution with 10 degrees of freedom,
the beta distribution with parameters $\alpha=9$ and $\beta=4$, and the Weibull distribution with shape parameter $k=2$ and scale parameter $\lambda=35$.
Other settings and the estimation procedure remain the same as in the previous study.
Finally, for estimators (\ref{ab}), (\ref{q1q3}), (\ref{abq}) and (\ref{shortcut}),
we simulate the data for each non-normal distribution with $500,000$ simulations,
and then report the natural logarithm of their RMSE values in Figure \ref{fig:distribution}
for $n$ up to 801, respectively.
From the numerical results, it is evident that
our new estimator is still able to provide a smaller RMSE value than the existing estimators in most settings.
This shows that our new estimator is quite robust to the violation of the normality assumption.
In particular, we note that estimator (\ref{ab}) performs even worse
when the sample size is large,  with the possible reason
that the minimum and maximum values are more likely to be the outliers
when they are simulated from heavy-tailed distributions.
In contrast,
our new estimator has a slowly increased RMSE as the sample size increases,
which also demonstrates that our new estimator has better asymptotic properties compared to the existing estimators.
Together with the comparison results in the previous study,
we conclude that our new estimator not only provides an optimal estimate of the sample SD for normal data,
but also performs favorably compared to the existing estimators for non-normal data.

\

\section{Conclusion}
\noindent
For clinical trials with continuous outcomes, the sample mean and SD are routinely reported in the literature.
While in some other studies, researchers may instead report the five-number summary including the sample median, the first and third quartiles, and the minimum and maximum values.
For these studies, when included in a meta-analysis, it is often desired to convert the five-number summary back to
the sample mean and standard deviation.
As reviewed in Section 2, a number of studies have emerged recently to solve this important problem under three common scenarios.
It is noted, however, that the existing methods, including Wan et al.\cite{wan2014estimating} and Bland,\cite{Bland2015Estimating} are still suboptimal for estimating the sample SD from the five-number summary.

To further advance the literature, we have proposed an improved estimator
for the sample SD by considering a smoothly weighted combination of two available estimators.
In addition, given that the analytical form of the optimal weight is complicated
and may not be readily accessible to practitioners,
we have also derived an approximation formula for the optimal weight,
and that yields a shortcut formula for the optimal estimation of the sample SD.
As confirmed by the theoretical and numerical results,
our new methods are able to dramatically improve the existing methods in the literature.
Together with Luo et al.,\cite{luo2018optimally}
we hence recommend practitioners to estimate the sample mean and SD from the five-number summary
by formulas (\ref{mabq1q3}) and (\ref{shortcut}), respectively.



To summarize, we have also provided the optimal estimators
of the sample mean and SD under the three common scenarios in Table \ref{summarytable}.
To be more specific, the optimal sample mean estimators under all three scenarios are from Luo et al.,\cite{luo2018optimally} the optimal sample SD estimators under scenarios $\mathcal{S}_1$ and $\mathcal{S}_2$ are from Wan et al.,\cite{wan2014estimating}
and the optimal sample SD estimator under scenario $\mathcal{S}_3$ is provided in (\ref{shortcut}) which makes
Table \ref{summarytable} a whole pie for data transformation from the five-number summary to the sample mean and SD.
To promote the practical use, we have also provided an Excel spreadsheet
to implement the optimal estimators in the Supplementary Material.
And more importantly, we have also incorporated the new estimator in this paper into our online calculator
at \url{http://www.math.hkbu.edu.hk/~tongt/papers/median2mean.html}.
According to Table \ref{summarytable}, if the five-number summary is reported for a certain study,
estimators (\ref{mabq1q3}) and (\ref{shortcut}) will be adopted to estimate the sample mean and SD, respectively.
Specifically, one can input the five-number summary and the sample size information
into the corresponding entries under scenario $\mathcal{S}_3$,
and then by clicking the ``Calculate" button,
the optimal estimates of the sample mean and SD will be automatically displayed in the result entries.

Finally, we note that all the estimators in Table \ref{summarytable} are established
under the normality assumption for the clinical trial data.
In practice, however, this normality assumption may not hold, in particular
when the five-number summary is reported rather than the sample mean and SD.
In view of this, researchers have also been proposing different approaches
for analyzing the studies reported with the five-number summary.
They include, for example, extending the data transformation methods
from normal data to non-normal data,\cite{kwon2015simulation}$^-$\cite{weir2018missing}
or developing new meta-analytical methods to directly synthesize the data with the five-number summary. \citealp{McGrath2018medians}$^,$\citealp{McGrath2019medians}
We note that the proposed method in this paper can also be readily extended to non-normal data,
yet further work is needed to assess the effectiveness of these new estimators when included in a meta-analysis.

\

\section{Theoretical results}
\noindent
We present the theoretical results of the proposed method in this section.
Specifically, we will have 2 theorems.
And to prove them, we need 2 lemmas as follows.

\begin{lemma}\label{lemma1}
Let $\Phi(x)$ and $\phi(x)$ be the cumulative distribution function and the probability density function of the standard normal distribution, respectively.
Let also $\Phi^{-1}$ be the inverse function of $\Phi$.
Then,
\begin{equation}\label{order1}
\lim \limits_{x \rightarrow 0^+} \dfrac{\Phi^{-1}(1-x)}{\sqrt{-2\ln(x)}}=1.
\end{equation}
\end{lemma}

\vskip 12pt
\noindent{\bf Proof}.
Since $\Phi^{-1}(1-x)$ and $\sqrt{-2\ln(x)}$ are both positive as $x \rightarrow 0^+$,
to prove formula (\ref{order1}) it is equivalent to showing that
\begin{equation*}
  \lim \limits_{x \rightarrow 0^+} \dfrac{\left[\Phi^{-1}(1-x)\right]^2}{-2\ln(x)}=1.
\end{equation*}
Let $y=\Phi^{-1}(1-x)$. Then, $x=1-\Phi(y)$. Noting that $y \rightarrow \infty$ as $x \rightarrow 0^+$,
by L'H$\hat{\rm o}$pital's rule, we have
\begin{align*}
     \lim \limits_{x \rightarrow 0^+} \dfrac{\left[\Phi^{-1}(1-x)\right]^2}{-2\ln(x)}
     = &\lim \limits_{y \rightarrow \infty} \dfrac{y^2}{-2{\rm ln}(1-\Phi(y))}\\
     = &\lim \limits_{y \rightarrow \infty} \dfrac{1-\Phi(y)}{y^{-1}\phi(y)}\\
     = &\lim \limits_{y \rightarrow \infty} \dfrac{\phi(y)}{\phi(y)+y^{-2}\phi(y)}\\
     = &1,
\end{align*}
where the second last equality follows by the Stein property of the standard normal distribution, that is, $d\phi(y)/dy = (-y)\phi(y)$.
\vskip 25pt

\begin{lemma}\label{lemma2}
Let $Z_{1} , Z_{2} , \ldots , Z_{n}$ follow the standard normal distribution with $Z_{(1)}\leq Z_{(2)} \leq \cdots \leq Z_{(n)}$ being the corresponding order statistics.
As $n\rightarrow \infty$, we have
\begin{equation*}\label{var_ab}
{\rm Var}(Z_{(n)}-Z_{(1)}) \approx \dfrac{\pi^2}{6\ln(n)},
\end{equation*}
\begin{equation}\label{var_q13}
  {\rm Var}(Z_{(3Q+1)}-Z_{(Q+1)}) \approx \dfrac{2.4758}{n},
\end{equation}
\begin{equation}\label{covar_abq}
  {\rm Cov}(Z_{(n)}-Z_{(1)},Z_{(3Q+1)}-Z_{(Q+1)}) = O\left(\dfrac{1}{\sqrt{n\ln(n)}}\right).
\end{equation}
\end{lemma}

\vskip 12pt
\noindent{\bf Proof}.
Since $Z_{(1)}$ and $Z_{(n)}$ are asymptotically independent (see Theorem 8.4.3 in Arnold et al.\cite{arnold2008first}),
then as $n \rightarrow \infty$, ${\rm Var}(Z_{(n)}-Z_{(1)}) \approx {\rm Var}(Z_{(1)})+{\rm Var}(Z_{(n)})$.
Note also that the limiting distribution of $Y_n=(Z_{(n)}-a_n)/b_n$ follows a Gumbel distribution as
$$\lim \limits_{n \rightarrow \infty}P\left(\dfrac{Z_{(n)}-a_n}{b_n}<x\right)=H(x)=\exp(-e^{-x}),$$
where $a_n=\left(2\ln(n)\right)^{1/2}-0.5(\ln(\ln(n))+\ln(4\pi))/(2\ln(n))^{1/2}$ and $b_n=(2\ln(n))^{-1/2}$.
In addition, for the Gumbel distribution,
the mean value is the Euler-Mascheroni constant $\gamma\approx0.5772$ and the variance is $\pi^2/6$.
Therefore, $E(Y_n)\approx \gamma$ and ${\rm Var}(Y_n)\approx \pi^2/6$ as $n\rightarrow \infty$,
and consequently, ${\rm Var}(Z_{(n)})=b_n^2{\rm Var}(Y_n)\approx \pi^2/(12\ln(n))$.
Further by symmetry, we have ${\rm Var}(Z_{(1)}) = {\rm Var}(Z_{(n)})$. Hence, as $n\rightarrow \infty$,
$$
{\rm Var}(Z_{(n)}-Z_{(1)}) \approx {\rm Var}(Z_{(1)}) + {\rm Var}(Z_{(n)}) \approx \dfrac{\pi^2}{6\ln(n)}.
$$

By Theorem 2 in Luo et al.,\cite{luo2018optimally} as $n\rightarrow \infty$,
${\rm Var}(Z_{(Q+1)}) = {\rm Var}(Z_{(3Q+1)}) \approx 1.8568/n$
and ${\rm Cov}(Z_{(Q+1)},Z_{(3Q+1)})\approx 0.6189/n$.
This shows that formula (\ref{var_q13}) holds.

By the Cauchy-Schwartz inequality, we have
\begin{equation*}
  {\rm Cov}(Z_{(1)},Z_{(Q+1)})  \leq\sqrt{{\rm Var}(Z_{(1)}){\rm Var}(Z_{(Q+1)})} = O\left(\dfrac{1}{\sqrt{n\ln(n)}}\right).
\end{equation*}
Similarly, we can show that ${\rm Cov}(Z_{(1)},Z_{(3Q+1)})$, ${\rm Cov}(Z_{(n)},Z_{(Q+1)})$
and ${\rm Cov}(Z_{(n)},Z_{(3Q+1)})$ are all of order $O(1/\sqrt{n\ln(n)})$.
Combining the above results, it is readily known that formula (\ref{covar_abq}) holds.
\vskip 25pt

\begin{theorem}\label{theorem1}
For the proposed estimator $S_w$ in formula (\ref{abq1q3}), we have the following properties.
\begin{enumerate}
\item[{\rm (i)}] $E(S_w)\approx \sigma$ for any weight $w$.
\item[{\rm (ii)}] The optimal weight is, approximately,\\
${w_{\rm opt}} = \dfrac{{\rm Var}(q_{3}-q_{1})/\eta^{2}-{\rm Cov}(b-a, q_{3}-q_{1})/(\xi\eta)}{{\rm Var}(b-a)/\xi^{2}+{\rm Var}(q_{3}-q_{1})/\eta^{2}-2{\rm Cov}(b-a, q_{3}-q_{1})/(\xi\eta)}$.
\item[{\rm(iii)}] $w_{\rm opt} = O(n^{-1/2}\ln (n))$ as $n \rightarrow \infty$.
\end{enumerate}
\end{theorem}

\vskip 12pt
\noindent{\bf Proof}.
(i) The expected value of the proposed estimator is
$$E(S_w)=wE\left(\dfrac{b-a}{\xi}\right)+(1-w)E\left(\dfrac{q_{3}-q_{1}}{\eta}\right).$$
Since $E\left((b-a)/\xi\right)\approx \sigma$ and $E\left((q_{3}-q_{1})/\eta\right)\approx \sigma$, we have $E(S_w)\approx \sigma$ for any weight $w$.

\vskip 12pt
(ii) By part (i), we have ${\rm Bias}(S_w)\approx0$.
Hence, to minimize the MSE of the estimator,
it is approximately equivalent to minimizing the variance of the estimator. Note that
\begin{equation}\nonumber
 {\rm Var}(S_w)
     =w^2{\rm Var}\left(\dfrac{b-a}{\xi}\right)+(1-w)^2{\rm Var}\left(\dfrac{q_{3}-q_{1}}{\eta}\right)+2w(1-w){\rm Cov}\left(\dfrac{b-a}{\xi},\dfrac{q_{3}-q_{1}}{\eta}\right).
\end{equation}
The first derivative of the variance with respect to $w$ is
\begin{align*}
  {d\over dw} {\rm Var}(S_w) = \ & 2w{\rm Var}\left(\dfrac{b-a}{\xi}\right)-2(1-w){\rm Var}\left(\dfrac{q_{3}-q_{1}}{\eta}\right) \\
            & +(2-4w){\rm Cov}\left(\dfrac{b-a}{\xi},\dfrac{q_{3}-q_{1}}{\eta}\right).
\end{align*}
Setting the first derivative equal to zero, we have
\begin{equation}\label{solution}
w= \dfrac{{\rm Var}(q_{3}-q_{1})/\eta^{2}-{\rm Cov}(b-a,q_{3}-q_{1})/(\xi\eta)}{{\rm Var}(b-a)/\xi^{2}+{\rm Var}(q_{3}-q_{1})/\eta^{2}-2{\rm Cov}(b-a,q_{3}-q_{1})/(\xi\eta)}.
\end{equation}
Also by the Cauchy-Schwartz inequality,
the second derivative is always non-negative,
$${d^2\over dw^2} {\rm Var}(S_w)= 2{\rm Var}\left(\dfrac{b-a}{\xi}\right)+2{\rm Var}\left(\dfrac{q_{3}-q_{1}}{\eta}\right)-4{\rm Cov}\left(\dfrac{b-a}{\xi},\dfrac{q_{3}-q_{1}}{\eta}\right)\geq0.$$
This shows that the weight in formula (\ref{solution}) is, approximately, the optimal weight of the estimator.

\vskip 12pt
(iii) Recall that, by formulas (\ref{approximation}) and (\ref{Jn}), the optimal weight can be rewritten as
\begin{equation}\label{w_new}
  w_{\rm opt} = \dfrac{{\rm Var}(Z_{(3Q+1)}-Z_{(Q+1)})/ \eta^2-{\rm Cov}(Z_{(n)}-Z_{(1)},Z_{(3Q+1)}-Z_{(Q+1)})/(\xi \eta)}{{\rm Var}[(Z_{(n)}-Z_{(1)})/ \xi-(Z_{(3Q+1)}-Z_{(Q+1)})/ \eta]},
\end{equation}
where $\xi = 2\Phi^{-1}[(n-0.375)/(n+0.25)]$ and $\eta = 2\Phi^{-1}[(0.75n-0.125)/(n+0.25)]$.
By Lemma \ref{lemma1}, as $n\rightarrow \infty$, we have
\begin{equation}\label{xi}
\xi = 2\Phi^{-1}\left(1-\dfrac{0.625}{n+0.25}\right) \approx 2\sqrt{-2{\rm ln}\left(\dfrac{0.625}{n+0.25}\right)}
= O(\sqrt{\ln(n)}).
\end{equation}
It is also clear that
\begin{equation}\label{eta}
\eta = 2\Phi^{-1}\left(\frac{0.75n-0.125}{n+0.25}\right) \approx 2\Phi^{-1}(0.75) = O(1).
\end{equation}
Then together with Lemma \ref{lemma2}, as $n\rightarrow \infty$,
the numerator of $w_{\rm opt}$ in formula (\ref{w_new}) is of order $O(n^{-1/2}\ln(n))$.
In addition, it can be shown that the denominator of $w_{\rm opt}$ is, approximately, $\pi^2[48\ln (n) \ln (n/0.625+0.4)]^{-1}$.
Finally, by combining the above results, we have $w_{\rm opt} = O(n^{-1/2}\ln (n))$ as $n\rightarrow \infty$.

\begin{theorem}\label{Rmse}
For the relative mean squared errors (RMSE) of the four differently weighted estimators,
as $n \rightarrow \infty$, we have
\begin{align*}
  {\rm RMSE}(S_1)  = \dfrac{{\rm MSE}(S_1)}{{\rm MSE}(S^{\rm Sam})} =O\left(\dfrac{n}{(\ln (n))^2}\right)
  ~~&~~ {\rm RMSE}(S_0)  = \dfrac{{\rm MSE}(S_0)}{{\rm MSE}(S^{\rm Sam})} \approx 2.721 \\
  {\rm RMSE}(S_{0.5})  = \dfrac{{\rm MSE}(S_{0.5})}{{\rm MSE}(S^{\rm Sam})} =O\left(\dfrac{n}{(\ln (n))^2}\right)  ~~&~~ {\rm RMSE}(S_{\tilde{w}_{\rm opt}}) = \dfrac{{\rm MSE}(S_{\tilde{w}_{\rm opt}})}{{\rm MSE}(S^{\rm Sam})} \approx 2.721,
\end{align*}
where $S_1$, $S_0$ and $S_{0.5}$ are Wan et al.'s estimators
(\ref{ab}), (\ref{q1q3}) and (\ref{abq}) respectively,
$S_{\tilde{w}_{\rm opt}}$ is our new estimator (\ref{shortcut}),
and $S^{\rm Sam}$ is the sample SD.
\end{theorem}

\vskip 12pt
\noindent{\bf Proof}.
By Holtzman,\cite{holtzman1950unbiased}
the expected value of the sample SD is $E(S^{\rm Sam})=c\sigma$, where $c=1-0.25/n+o(1/n)$.
Note also that $(S^{\rm Sam})^2$ is the sample variance and so is an unbiased estimator for $\sigma^2$. Then,
\begin{equation}\label{mse_sam}
{\rm MSE}(S^{\rm Sam}) =  E(S^{\rm Sam})^2-2\sigma E(S^{\rm Sam})+\sigma^2= 2\sigma^2-2c\sigma^2 =  0.5 \sigma^2/n+o(1/n).
\end{equation}

For the existing estimators, by Wan et al.,\cite{wan2014estimating}
we have $E(S_w) \approx \sigma$ for $w=0, 0.5\ {\rm and} \ 1$.
Further by Lemma \ref{lemma2} and formulas (\ref{xi})-(\ref{eta}), as $n\rightarrow \infty$,
\begin{equation*}\label{var_s1}
  {\rm MSE}(S_1) \approx \dfrac{{\rm Var}(Z_{(n)}-Z_{(1)})}{4\xi^2}\sigma^2=O\left(\dfrac{1}{(\ln (n))^2}\right),
\end{equation*}
\begin{equation*}\label{var_s2}
  {\rm MSE}(S_0) \approx \dfrac{{\rm Var}(Z_{(3Q+1)}-Z_{(Q+1)})}{4\eta^2}\sigma^2
  \approx \dfrac{2.4758\sigma^2/n}{4[\Phi^{-1}(0.75)]^2}+o\left(\dfrac{1}{n}\right) \approx \dfrac{1.3605\sigma^2}{n}+o\left(\dfrac{1}{n}\right),
\end{equation*}
\begin{align*}\label{var_ex}
  {\rm MSE}(S_{0.5}) \approx & \ \sigma^2\bigg[\dfrac{{\rm Var}(Z_{(n)}-Z_{(1)})}{4\xi^2}+\dfrac{{\rm Var}(Z_{(3Q+1)}-Z_{(Q+1)})}{4\eta^2}\\
  &+\dfrac{{\rm Cov}(Z_{(n)}-Z_{(1)},Z_{(3Q+1)}-Z_{(Q+1)})}{2\xi\eta}\bigg]\\
  =&\ O\left(\dfrac{1}{(\ln (n))^2}\right)+O\left(\dfrac{1}{n}\right)+O\left(\dfrac{1}{n^{1/2}\ln (n)}\right)
  \\
  =&\ O\left(\dfrac{1}{(\ln (n))^2}\right),
\end{align*}
where $\Phi^{-1}(0.75)\approx 0.6745$.
Then together with formula (\ref{mse_sam}), we have ${\rm RMSE}(S_1) = O(n/(\ln (n))^2)$,
${\rm RMSE}(S_0) = 2.721$
and ${\rm RMSE}(S_{0.5}) = O(n/(\ln (n))^2)$ as $n \rightarrow \infty$.

For our new estimator, by Theorem \ref{theorem1}, we have $E(S_{\tilde{w}_{\rm opt}})\approx \sigma$.
Note also that $\theta_1=(1+0.07n^{0.6})\xi$ and $\theta_2=[1+n^{-0.6}/0.07]\eta$.
Then by Lemma \ref{lemma2} and formulas (\ref{xi})-(\ref{eta}), as $n\to\infty$,
\begin{align*}
  {\rm MSE}(S_{\tilde{w}_{\rm opt}})
  \approx & \ \sigma^2\bigg[\dfrac{{\rm Var}(Z_{(n)}-Z_{(1)})}{(1+0.07n^{0.6})^2\xi^2}+\dfrac{{\rm Var}(Z_{(3Q+1)}-Z_{(Q+1)})}{[1+1/(0.07n^{0.6})]^2\eta^2}\\
  &+\dfrac{2{\rm Cov}\left(Z_{(n)}-Z_{(1)},Z_{(3Q+1)}-Z_{(Q+1)}\right)}{(1+0.07n^{0.6})[1+1/(0.07n^{0.6})]\xi\eta}\bigg] \\
  \approx & \ O\left(\dfrac{1}{n^{1.2}(\ln (n))^2}\right)+\dfrac{2.4758\sigma^2/n}{4[\Phi^{-1}(0.75)]^2}+O\left(\dfrac{1}{n^{1.1}\ln (n)}\right) \\
  = & \ \dfrac{1.3605\sigma^2}{n}+o\left(\dfrac{1}{n}\right).
\end{align*}
Finally, together with formula (\ref{mse_sam}), we have ${\rm RMSE}(S_{\tilde{w}_{\rm opt}})\approx 2.721$ as $n \rightarrow \infty$.

\

\section*{Highlights}
\begin{itemize}
\item What is already known:

A few methods for data transformation from the sample median, the minimum and maximum values, and/or the first and third quartiles to the sample mean and standard deviation have been proposed in the literature,
in which Luo et al.\cite{luo2018optimally} provided the optimal sample mean estimators under three common scenarios $\mathcal{S}_1$, $\mathcal{S}_2$ and $\mathcal{S}_3$ and Wan et al.\cite{wan2014estimating} provided the optimal estimators of the sample standard deviation under scenarios $\mathcal{S}_1$ and $\mathcal{S}_2$.

\item What is new:

We propose a smoothly weighted estimator for the sample standard deviation under scenario $\mathcal{S}_3$
and also establish a shortcut formula of our optimal estimator.
For practical use, we also provide an Excel spreadsheet and an online calculator
to implement the optimal estimation of the sample mean and standard deviation from the five-number summary.

\item Potential impact for RSM readers:

Together with Luo et al.\cite{luo2018optimally} and Wan et al.,\cite{wan2014estimating}
our optimal estimators are capable to serve as
``rules of thumb" for estimating the sample mean and standard deviation from the reported quantiles.
They make a solid contribution to meta-analysis and will have the potential to be widely used.
\end{itemize}

\vskip -36pt
\section*{Acknowledgements}
\noindent
The authors thank the editor, the associate editor, and
the reviewer for their constructive comments that have led to
a significant improvement of the paper.
Lu Lin's research was supported in part by the National Natural
Science Foundation of China (No. 11971265).
Haitao Chu's research was supported in part by the NIH National Library of Medicine (No. R01LM012982).
Tiejun Tong's research was supported in part by the Initiation Grant for Faculty
Niche Research Areas (No. RC-IG-FNRA/17-18/13) and the Century Club Sponsorship
Scheme of Hong Kong Baptist University, the National Natural
Science Foundation of China (No. 11671338), and the General
Research Fund (No. HKBU12303918).
\vskip -48pt
\section*{Conflict of Interest}
\noindent
The authors reported no conflict of interest.
\section*{Data Availability Statement}
\noindent
The numerical values of $\theta_1$ and $\theta_2$ in the shortcut formula (\ref{shortcut}) for $Q$ up to 100
are available in Table \ref{coefficients}.
\section*{Supplementary Material}
\noindent
An Excel spreadsheet for estimating the sample mean and standard deviation from the five-number summary.

\newpage

\newpage

\newpage
\begin{table}[H]\normalsize
\setlength{\abovecaptionskip}{0pt}%
\setlength{\belowcaptionskip}{6pt}%

\caption{Values of $\theta_1$ and $\theta_2$ in estimator (\ref{shortcut}) for $1\leq Q\leq100$, where $n=4Q+1$.}\label{coefficients}
\centering
\begin{tabular}{ccc|ccc|ccc|ccc}
\hline
$Q$ &$\theta_1$	&$\theta_2$  &$Q$	   &$\theta_1$	&$\theta_2$  &$Q$	   &$\theta_1$	&$\theta_2$  &$Q$	   &$\theta_1$	&$\theta_2$\\
\hline
1	&2.793	&6.403 &26	&10.780	&2.495	&51	&14.848	&2.124		
&76	&18.178	&1.962\\
2	&3.770	&5.514	&27	&10.967	&2.470	&52	&14.992	&2.115	
&77	&18.302	&1.958	\\
3	&4.437	&4.895	&28	&11.151	&2.447	&53	&15.135	&2.107
&78	&18.425	&1.953	\\
4	&4.969	&4.466	&29	&11.333	&2.425	&54	&15.276	&2.099
&79	&18.548	&1.949	\\
5	&5.423	&4.150	&30	&11.512	&2.404	&55	&15.417	&2.091
&80	&18.670	&1.944 \\
6	&5.826	&3.906	&31	&11.689	&2.385	&56	&15.557	&2.083	
&81	&18.791	&1.940\\
7	&6.192	&3.712	&32	&11.863	&2.366	&57	&15.695	&2.075
&82	&18.912	&1.935	\\
8	&6.531	&3.553	&33	&12.036	&2.348	&58	&15.833	&2.068
&83	&19.033	&1.931	\\
9	&6.848	&3.419	&34	&12.206	&2.331	&59	&15.970	&2.061
&84	&19.152	&1.927	\\
10	&7.147	&3.305	&35	&12.374	&2.314	&60	&16.106	&2.054	
&85	&19.272	&1.923  \\
11	&7.432	&3.206	&36	&12.540	&2.299	&61	&16.241	&2.047
&86	&19.391	&1.919	\\
12	&7.704	&3.120	&37	&12.705	&2.284	&62	&16.375	&2.040
&87	&19.509	&1.915	\\
13	&7.966	&3.044	&38	&12.867	&2.269	&63	&16.509	&2.034
&88	&19.627	&1.912	\\
14	&8.218	&2.975	&39	&13.028	&2.256	&64	&16.642	&2.028
&89	&19.744	&1.908	\\
15	&8.462	&2.914	&40	&13.188	&2.242	&65	&16.774	&2.021
&90	&19.861	&1.904	\\
16	&8.699	&2.859	&41	&13.345	&2.230  &66	&16.905	&2.015
&91	&19.977	&1.901	\\
17	&8.929	&2.808	&42	&13.502	&2.217	&67	&17.035	&2.010
&92	&20.093	&1.897	\\
18	&9.153	&2.762	&43	&13.656	&2.205  &68	&17.165	&2.004
&93	&20.208	&1.893	\\
19	&9.371	&2.719	&44	&13.810	&2.194	&69	&17.294	&1.998
&94	&20.323	&1.890	\\
20	&9.585	&2.680	&45	&13.962	&2.183	&70	&17.422	&1.993
&95	&20.438	&1.887	\\
21	&9.793	&2.644  &46	&14.113	&2.172	&71	&17.550	&1.987
&96	&20.552	&1.883	\\
22	&9.998	&2.610	&47	&14.262	&2.162	&72	&17.677	&1.982
&97	&20.666	&1.880	\\
23	&10.199	&2.578	&48	&14.410	&2.152	&73	&17.803	&1.977
&98	&20.779	&1.877	\\
24	&10.396	&2.549	&49	&14.558	&2.143	&74	&17.929	&1.972
&99	&20.892	&1.874	\\
25	&10.589	&2.521	&50	&14.703	&2.133  &75	&18.054	&1.967
&100	&21.004	&1.871	\\

\hline
\end{tabular}
\end{table}

\newpage
\begin{table}[H]\normalsize
\setlength{\abovecaptionskip}{0pt}%
\setlength{\belowcaptionskip}{6pt}%

\caption{Summary table of the optimal estimators of the sample mean and SD under the three common scenarios,
where $w_1=4/(4+n^{0.75})$ and $w_2=0.7+0.39/n$.}\label{summarytable}
\centering
\begin{tabular}{c|cc}
\hline
~Scenario        ~~&~~Sample mean estimator       ~~~~&~~~~Sample SD estimator~ \\
\hline
~$\mathcal{S}_1=\{a,m,b;n\}$ ~~&~~ $w_1(a+b)/2+(1-w_1)m$      ~~~~&~~~~formula (\ref{ab})~\\
~$\mathcal{S}_2=\{q_1,m,q_3;n\}$ ~~&~~ $w_2(q_1+q_3)/2+(1-w_2)m$  ~~~~&~~~~formula (\ref{q1q3})~\\
~$\mathcal{S}_3=\{a,q_1,m,q_3,b;n\}$ ~~&~~ formula (\ref{mabq1q3})    ~~~~&~~~~formula (\ref{shortcut})~\\
\hline
\end{tabular}
\end{table}

\end{document}